\documentclass[a4paper]{article}
\usepackage{basicstyle}
\usepackage{natbib}
\usepackage{amsthm}
\usepackage[framemethod=TikZ]{mdframed} % framed theorems

\usepackage{hyperref}
\usepackage{tikz}
\usepackage{glossaries}

\theoremstyle{definition}
\newmdtheoremenv{example}{Example}
\AfterEndEnvironment{example}{\noindent\ignorespaces}

\newglossaryentry{feature}
{
    name = {feature},
    description = {}
}

\newglossaryentry{targetvariable}
{
    name = {target variable},
    description = {}
}

\newglossaryentry{performancemetric}
{
    name = {performance metric},
    description = {}
}

\newglossaryentry{machinelearning}{
    name = {machine learning},
    description = {A subfield of artificial intelligence that  to build models that allow a computer to perform a particular task by learning from experience, without being explicitly programmed}
}

\newglossaryentry{instance}{
    name = {instance},
    description = {An entry of a data set}
}

\newglossaryentry{supervisedlearning}{
    name = {supervised learning},
    description = {A category of machine learning that requires explicit }
}

\newglossaryentry{classifier}{
    name = {classifier},
    description = {A machine learning model designed for \gls{classification}}
}

\newglossaryentry{unsupervisedlearning}{
    name = {unsupervised learning},
    description = {}
}

\newglossaryentry{reinforcementlearning}{
    name = {reinforcement learning},
    description = {}
}

\newglossaryentry{regression}{
    name = {regression},
    description = {Regression is a type of \gls{supervisedlearning} that aims at predicting a real-valued target variable}
}

\newglossaryentry{lossfunction}{
    name = {loss function},
    description = {}
}

\newglossaryentry{classification}{
    name = {classification},
    description = {Classification is a type of \gls{supervisedlearning} that aims at predicting discrete \glspl{class}}
}

\newglossaryentry{class}{
    name = {class},
    description = {}
}

\newglossaryentry{imbalanced}{
    name = {imbalanced},
    description = {A data set is considered imbalanced if the number of \glspl{instance} per \gls{class} differs greatly across classes}
}

\newglossaryentry{binaryclassification}{
    name = {binary classification},
    description = {A type of classification where the \gls{targetvariable} consists of two \glspl{class}}
}

\newglossaryentry{negativeclass}{
    name = {negative class},
    description = { See also \gls{positiveclass}}
}

\newglossaryentry{positiveclass}{
    name = {positive class},
    description = { See also \gls{negativeclass}}
}

\newglossaryentry{accuracy}{
    name = {accuracy},
    description = {}
}

\newglossaryentry{precision}{
    name = {precision},
    description = {}
}

\newglossaryentry{recall}{
    name = {recall},
    description = {}
}

\newglossaryentry{f1score}{
    name = {F1-score},
    description = {A predictive performance metric that is computed as the harmonic mean of \gls{precision} and \gls{recall}}
}

\newglossaryentry{confusionmatrix}{
    name = {confusion matrix},
    description = {A table that visualizes the predictive performance of a classification algorithm as the number of instances in some \gls{class} $i$ that were predicted to be class $j$}
}

\newglossaryentry{truepositive}{
    name = {true positive},
    description = {}
}

\newglossaryentry{falsepositive}{
    name = {false positive},
    description = {}
}

\newglossaryentry{truenegative}{
    name = {true negative},
    description = {}
}

\newglossaryentry{falsenegative}{
    name = {false negative},
    description = {}
}

\newglossaryentry{truepositiverate}{
    name = {true positive rate},
    description = {}
}

\newglossaryentry{truenegativerate}{
    name = {true negative rate},
    description = {}
}
\newglossaryentry{falsepositiverate}{
    name = {false positive rate},
    description = {}
}

\newglossaryentry{falsenegativerate}{
    name = {false negative rate},
    description = {}
}

\newglossaryentry{roccurve}{
    name = {ROC curve},
    description = {The Receiver Operating Characteristic (ROC) curve is a plot that sets out the \gls{truepositiverate} of a classifier against the \gls{falsepositiverate} for different decision thresholds}
}

\newglossaryentry{precisionrecallcurve}{
    name = {precision-recall curve},
    description = {The precision-recall curve is a plot that sets out the \gls{precision} of a classifier against it's \gls{recall} for different decision thresholds}
}

\newglossaryentry{hyperparameter}{
    name = {hyperparameter},
    description = {Parameters of \gls{machinelearning} algorithms that control the learning process}
}

\newglossaryentry{decisionthreshold}{
    name = {decision threshold},
    description = {The decision threshold is the cut-off value }
}

\newglossaryentry{auc}{
    name = {AUC},
    description = {Area under the ROC-curve.}
}

\newglossaryentry{averageprecision}{
    name = {average precision},
    description = {}
}

\newglossaryentry{confidencescore}
{
    name = {confidence score},
    description = {}
}

\newglossaryentry{costsensitivelearning}
{
    name = {cost sensitive learning},
    description = {}
}

\newglossaryentry{calibration}
{
    name = {calibration},
    description = {}
}

\newglossaryentry{overfitting}
{
    name = {overfitting},
    description = {A situation in which a machine learning model has learned artefacts of the training data that do not generalize to new unseen data. See also \gls{underfitting}}
}

\newglossaryentry{underfitting}
{
    name={underfitting},
    description={A situation in which a machine learning model is not able to adequately capture the underlying structure of the data. See also \gls{overfitting}}
}

\newglossaryentry{trainingset}
{
    name = {training set},
    description = {The data set used to fit a machine learning model. See also \gls{validationset} and \gls{testset}}
}

\newglossaryentry{validationset}
{
    name = {validation set},
    description = {The data set used to estimate a machine learning model's performance during model selection. See also \gls{trainingset} and \gls{testset}}
}

\newglossaryentry{testset}
{
    name = {test set},
    description = {The data set used to estimate the predictive performance of the final model that was selected during model selection. See also \gls{trainingset} and \gls{validationset}}
}

\newglossaryentry{crossvalidation}
{
    name = {k-fold cross-validation},
    description = {A procedure used to estimate the performance of a machine learning algorithm with particular hyperparameter settings on a data set, in which the data set is split into $k$ folds. For each fold, a machine learning model is trained on the remaining $k-1$ folds and tested on the current fold. The overall performance of the algorithm is estimated based on the performance on all folds}
}

\newglossaryentry{repeatedcrossvalidation}
{
    name = {repeated k-fold cross-validation},
    description = {A variant of \gls{crossvalidation} in which the cross-validation procedure is repeated multiple times, shuffling the data into different folds for each repeat},
    parent={crossvalidation}
}

\newglossaryentry{nestedcrossvalidation}
{
    name = {nested k-fold cross-validation},
    description = {A variant of \gls{crossvalidation} in which the cross-validation procedure is nested. In the outer loop, the data set is split into folds used to determine the training/validation and test set. In the inner loop, the training/validation set of each test fold of the outer loop is further split into folds used for training and validation},
    parent={crossvalidation}
}

\newglossaryentry{gridsearch}{
    name = {grid search}, 
    description = {A \gls{hyperparameter} tuning approach which consists of an exhaustive search of hyperparameter settings over a predefined search space. See also \gls{randomsearch}}
}
    
\newglossaryentry{randomsearch}{
    name = {random search search}, 
    description = {A \gls{hyperparameter} tuning approach in which hyperparameter settings are sampled randomly from a predefined search space. See also \gls{gridsearch}}
}

\newglossaryentry{moralvalue}
{
    name = {moral value},
    description = {A conviction or matter that people feel should be strived for in general and not just for themselves to be able to lead a good life or to realize a just society \citep{Poel2011}}
}

\newglossaryentry{moralvirtue}
{
    name = {moral virtue},
    description = {A character trait that make somebody a good person or allows people to lead good lives \citep{Poel2011}}
}

\newglossaryentry{moralnorm}
{
    name = {moral norm},
    description={A rule that prescribe what actions are required, permitted, or forbidden \citet{Poel2011}}
}

\newglossaryentry{moralprinciple}
{
    name = {moral principle},
    description = {} 
}

\newglossaryentry{morality}
{
    name = {morality},
    description = {The totality of opinions, decisions, and actions with which people express, individually or collectively, what they think is good or right \citep{Poel2011}}
}

\newglossaryentry{consequentialism}
{
    name = {consequentialism},
    description = {An ethical framework from Western philosophy that poses that the consequences of an act determine whether it is morally acceptable. See also \gls{utilitarianism}}
}

\newglossaryentry{utilitarianism}
{
    name = {utilitarianism},
    description = {An ethical framework from Western philosophy that poses that an act is only morally acceptable if it can be reasonably expected to bring the greatest happiness to the greatest number of people. It is a form of \gls{consequentialism}}
}

\newglossaryentry{deontology}
{
    name = {deontology},
    description = {An ethical framework from Western philosophy that poses that an act is only morally acceptable if you believe that it should be a universal law}
}

\newglossaryentry{moralproblem}
{
    name = {moral problem},
    description = {A problem where two or more moral values or norms cannot be fully realized at the same time \citep{Poel2011}}
}

\newglossaryentry{commonsensemethod}
{
    name = {common sense method},
    description = {The common sense method weights the available options for actions in the light of the relevant (moral) values. \citep{Poel2011}}
}

\newglossaryentry{abstractiontrap}
{   
    name={abstraction trap},
    description={A failure to consider how social context is interlaced with technology due to abstracting away context surrounding a machine learning model and its inputs and outputs \citep{Selbst2019}}
}

\newglossaryentry{solutionismtrap}
{   
    name=solutionism trap,
    description={A failure to consider that machine learning might not be the best solution to a problems \citep{Selbst2019}. See als \gls{solutionism}},
    parent={abstractiontrap}
}

\newglossaryentry{solutionism}
{
    name=solutionsm,
    description={The belief that every problem can be solved using technology. See also \gls{solutionismtrap}}
}

\newglossaryentry{rippleffecttrap}
{   
    name=ripple effect trap,
    description={A failure to consider the ways in which the introduction of a machine learning model may affect behavior of other actors in the system and change the social context \citep{Selbst2019}},
    parent={abstractiontrap}
}

\newglossaryentry{formalismtrap}
{   
    name=formalism trap,
    description={A failure to take into account sociotechnical context when formalizing the problem \citep{Selbst2019}. Closely related to \gls{constructvaliditybias}},
    parent={abstractiontrap}
}

\newglossaryentry{portabilitytrap}
{   
    name=portability trap,
    description={Taking an existing solution and applying it in a different situation without taking into account the differences between the two contexts \citep{Selbst2019}},
    parent={abstractiontrap}
}

\newglossaryentry{framingtrap}
{   
    name=framing trap,
    description={\citep{Selbst2019}},
    parent={abstractiontrap}
}

\newglossaryentry{hazard}
{
    name=hazard,
    description={A potential source of a \gls{harm}}
}

\newglossaryentry{risk}
{
    name=risk,
    description={The }
}

\newglossaryentry{harm}
{
    name=harm,
    description={Physical or psychological injury or damage}
}

\newglossaryentry{allocationharm}
{   
    name=allocation harm,
    description={Disproportionate allocation of opportunities, resources, or information},
    parent={harm}
}

\newglossaryentry{qualityofserviceharm}
{   
    name=quality-of-service harm,
    description={Disproportionate failure of a system for certain (groups of) people},
    parent={harm}
}

\newglossaryentry{stereotypingharm}
{   
    name=stereotyping harm,
    description={Reinforcement of undesirable and unfair societal stereotypes},
    parent={harm}
}

\newglossaryentry{denigrationharm}
{   
    name=denigration harm,
    description={Actively derogatory or offensive behavior against certain (groups of) people},
    parent={harm}
}

\newglossaryentry{representationharm}
{   
    name=representation harm,
    description={Over- or under-representation of certain groups of people},
    parent={harm}
}

\newglossaryentry{proceduralharm}
{
    name = procedural harm, 
    description = {Decisions are made in a way that violates social norms},
    parent = {harm}
}

\newglossaryentry{algorithmicfairness}
{   
    name=algorithmic fairness,
    description={The idea that algorithmic systems should behave or treat people fairly, i.e., without discrimination on the grounds of \glspl{sensitivecharacteristic} such as age, sex, disability, ethnic or racial origin, religion or belief, or sexual orientation. Synonym of \gls{fairness}}
}

\newglossaryentry{fairness}
{
    name=fairness,
    description={See \gls{algorithmicfairness}}
}

\newglossaryentry{directdiscrimination}
{
    name=direct discrimination,
    description={A legal concept that considers a case where (groups of) individuals are treated less
favorably based directly on their membership of a protected group}
}

\newglossaryentry{indirectdiscrimination}
{
    name=indirect discrimination,
    description={A legal concept that considers a case where (groups of) individuals are treated less
favorably based on rules that seem neutral, but, as a side effect, disadvantage a protected group}
}

\newglossaryentry{sensitivecharacteristic}
{
    name = sensitive characteristic,
    description={A characteristic of an individual such that any decisions based on this characteristic are considered undesirable from an ethical or legal point of view. Examples are age, sex, disability, ethnic or racial origin, religion or belief, or sexual orientation. See also \gls{sensitivefeature}}
}

\newglossaryentry{sensitivefeature}
{
    name = sensitive feature,
    description={A feature of a data set that represents a \gls{sensitivecharacteristic}}
}

\newglossaryentry{counterfactualfairness}
{
    name=counterfactual fairness,
    description={A notion of fairness that requires that the treatment of an individual in the actual world is the same as the treatment of the individual in a counterfactual world where the individual belonged to a different sensitive group.},
    parent={algorithmicfairness}
}

\newglossaryentry{individualfairness}
{
    name=individual fairness,
    description={A notion of fairness that requires similar cases to be treated similarly, irrespective of sensitive group membership},
    parent={algorithmicfairness}
}

\newglossaryentry{statisticalfairness}
{
    name=statistical fairness,
    description={See \gls{groupfairness}},
    parent={algorithmicfairness}
}

\newglossaryentry{groupfairness}
{
    name=group fairness,
    description={A notion of fairness that requires group statistics (e.g., selection rate, false positive rate, or calibration) to be equal across (sub)groups defined by sensitive characteristics. Synonym of \gls{statisticalfairness}},
    parent={algorithmicfairness}
}

\newglossaryentry{conditionalgroupfairness}
{
    name=conditional group fairness,
    description={A variant of \gls{groupfairness} that allows for differences between groups, if these differences are explained by a legitimate feature that can be justified by ethics and/or law},
    parent={algorithmicfairness}
}

\newglossaryentry{proceduralfairness}
{
    name = procedural fairness,
    description={A notion of fairness that requires that the predictive model uses features in a way that does not violate social norms.},
    parent={algorithmicfairness}}
    
\newglossaryentry{simpsonsparadox}{
    name=Simpson's paradox,
    description = {If a correlation occurs in several different groups, it may disappear or even reverse when the groups are aggregated}
}

\newglossaryentry{situationtesting}{
    name = situation testing,
    description = {A traditional discrimination auditing approach in which pairs of individuals that are equal, except for their membership of a protected group, are put in the same situation. If the member of a protected group is treated less favorably, this is regarded discrimination}
}

\newglossaryentry{abstraction}{
    name = abstraction,
    description = {}
}

\newglossaryentry{redlining}{
    name = redlining,
    description = {A practice in the United States where people were systematically denied services based on their postal code, which is highly correlated with race, indirectly discriminating predominantly Black neighborhoods}
}

\newglossaryentry{constructvalidity}{
    name = construct validity,
    description = {The extent to which a measurement measures the intended construct}
}

\newglossaryentry{measurementmodel}{
    name = measurement model,
    description = {A model that specifies how a set of variables can be combined to represent an \gls{unobservableconstruct}. A \gls{proxyvariable} is an example of a simple measurement model with only one variable}
}
    
\newglossaryentry{unobservableconstruct}{
    name = unobservable construct,
    description = {An abstract concept or idea that is impossible to measure directly. An example is socio-economic status}}

\newglossaryentry{optimismbias}{
    name = optimism bias,
    description = {A \gls{cognitivebias} that causes people to overestimate the likelihood of positive events and underestimate the likelihood of negative events}
}

\newglossaryentry{confirmationbias}{
    name = confirmation bias,
    description ={A tendency to put attention towards information that confirms one's own beliefs}
}

\newglossaryentry{conceptdrift}{
    name = concept drift,
    description = {A situation in which the statistical properties of the target variable change over time}
}

\newglossaryentry{approximatefairness}
{
    name=approximate fairness,
    description={A relaxation of a fairness criterion that allows some deviation of the criterion (as opposed to exact equality) between groups.},
    parent={algorithmicfairness}
}

\newglossaryentry{fairnessmetric}
{
    name=fairness metric,
    description={A metric that quantifies to what extent predictions deviate from a \gls{fairnesscriterion}.},
}

\newglossaryentry{fairnesscriterion}
{
    name=fairness criterion,
    description={A criterion that formalizes a condition for fairness, such as equal selection rates between groups.},
}

\newglossaryentry{independencecriterion}
{
    name=independence criterion,
    description={A category of \glspl{fairnesscriterion} that requires the sensitive features $A$ to be statistically independent of the predictions $\hat{Y}$.; i.e., $A \perp \hat{Y}$}
}

\newglossaryentry{separationcriterion}
{
    name=separation criterion,
    description={A category of \glspl{fairnesscriterion} that requires the predictions $\hat{Y}$ to be statistically independent of the sensitive features $\hat{A}$, conditional on the ground truth labels ${Y}$; i.e., $\hat{Y} \perp {A} \mid Y$},
}

\newglossaryentry{sufficiencycriterion}
{
    name=sufficiency criterion,
    description={A category of \glspl{fairnesscriterion} that requires the ground truth labels $Y$ to be independent of the sensitive features $A$, conditional on the predictions $\hat{Y}$; i.e., $Y \perp {A} \mid \hat{Y}$},
}

\newglossaryentry{demographicparity}
{
    name=demographic parity,
    description={A \gls{fairnesscriterion}},
}

\newglossaryentry{conditionaldemographicparity}
{
    name=conditional demographic parity,
    description={A \gls{fairnesscriterion}},
}

\newglossaryentry{equalodds}
{
    name=equal odds,
    description={A \gls{fairnesscriterion}},
}

\newglossaryentry{equalopportunity}
{
    name=equal opportunity,
    description={A \gls{fairnesscriterion}},
}

\newglossaryentry{equalcalibration}
{
    name=equal calibration,
    description={A \gls{fairnesscriterion}},
}

\newglossaryentry{predictiveparity}
{
    name=predictive parity,
    description={A \gls{fairnesscriterion}},
}

\newglossaryentry{reinforcingfeedbackloop}
{
    name=reinforcing feedback loop,
    description={Feedback mechanisms that amplify an effect. In the context of \gls{algorithmicfairness}, it refers to the amplification of existing biases when new data is collected based on the output of a biased model. See also \gls{selectionbias}}
}

\newglossaryentry{bias}
{
    name=bias,
    description={A systematic and disproportionate tendency towards something}
}

\newglossaryentry{socialbias}
{
    name=social bias,
    description={A prejudice for or against a person or a group},
    parent={bias}
}

\newglossaryentry{cognitivebias}
{
    name=cognitive bias,
    description={A systematic error in rational thinking that can affect judgement and decision-making},
    parent={bias}
}

\newglossaryentry{statisticalbias}
{
    name=statistical bias,
    description={A systematic error in the estimation of parameters or variables},
    parent={bias}
}

\newglossaryentry{optimisimbias}
{
    name=optimism bias,
    description={A \gls{cognitivebias} that causes people to overestimate the likelihood of positive ivents and underestimate the likelihood of negative events},
    parent={bias}
}

\newglossaryentry{constructvaliditybias}
{
    name=construct validity bias,
    description={A \gls{statisticalbias} that occurs when a feature or target variable does not accurately measure the construct it was designed to measure},
    parent={bias}
}

\newglossaryentry{historicalbias}
{
    name=historical bias,
    description={A \gls{socialbias} that is encoded in the data through biased human decision-making or structural biases embedded in society},
    parent={bias}
}

\newglossaryentry{representationbias}
{
    name=representation bias,
    description={A bias that occurs when some groups, defined by sensitive characteristics, are underrepresented in the data},
    parent={bias}
}

\newglossaryentry{selectionbias}
{
    name=selection bias,
    description={A \gls{statisticalbias} that occurs when data collection or selection procedures results in a non-random sample of the population},
    parent={bias}
}

\newglossaryentry{measurementbias}
{
    name=measurement bias,
    description={A \gls{statisticalbias} that occurs when data collection practices result in systematic errors in data. Can cause \gls{constructvaliditybias},
    parent={bias}}
}

\newglossaryentry{aggregationbias}
{
    name = aggregation bias,
    description = {A bias that occurs when a single machine learning model is used for groups that have distinct data  distributions, resulting in inaccurate predictions for (some) groups},
    parent={bias}
}

\newglossaryentry{evaluationbias}
{
    name = {evaluation bias},
    description = {A bias that occurs through the use of performance metrics and procedures that are not appropriate for the context in which the model will be used},
    parent={bias}
}

\newglossaryentry{popularitybias}
{
    name = {popularity bias},
    description = {A bias that occurs in recommender systems when highly ranked items are selected more often, increasing their ranking even further, while disregarding less popular items that may be equally valuable},
    parent={bias}
}

\newglossaryentry{omittedvariablebias}{
    name = {omitted variable bias},
    description = {A \gls{statisticalbias} that occurs when a model leaves out one or more relevant variables},
    parent={bias}
}

\newglossaryentry{explainableml}
{   
    name=explainable machine learning,
    description={A branch of machine learning that studies explaining machine learning models' inner workings or predictions in a way that is understandable to humans. Sometimes used interchangeably with \gls{interpretableml}}
}

\newglossaryentry{interpretableml}
{   
    name=interpretable machine learning,
    description={A branch of machine learning that studies explaining machine learning models' inner workings or predictions in a way that is understandable to humans. Sometimes used interchangeably with \gls{explainableml}}
}

\newglossaryentry{proxyvariable}
{
    name=proxy variable,
    description={A variable that is highly associated with another variable. Proxy variables often act as a simple measurement model for an immeasurable construct or a variable that is otherwise hard to measure directly}
}

\newglossaryentry{algorithmicaccountability}
{
    name = algorithmic accountability,
    description = {}}

\makeglossaries

\begin{document}
%
% ------------ Title ------------ %
\title{An Introduction to Algorithmic Fairness}
% \subtitle{Lecture Notes 2IX30}
\author{
    Hilde J.P. Weerts\\
    Eindhoven University of Technology\\ 
    \href{mailto:h.j.p.weerts@tue.nl}{h.j.p.weerts@tue.nl}
}
\maketitle
%
% ------------ Body ------------ %

\begin{abstract}
In recent years, there has been an increasing awareness of both the public and scientific community that algorithmic systems can reproduce, amplify, or even introduce unfairness in our societies. These lecture notes provide an introduction to some of the core concepts in algorithmic fairness research. We list different types of fairness-related harms, explain two main notions of algorithmic fairness, and map the biases that underlie these harms upon the machine learning development process.
\end{abstract}
In recent years, there has been an increasing awareness of both the public and scientific community that algorithmic systems can reproduce, amplify, or even introduce unfairness in our societies. From automated resume screening tools that favor men over women to facial recognition systems that disproportionately fail for darker-skinned women compared to white men \citep{buolamwini2018}. In these lecture notes, we will introduce several ways in which machine learning can result in discrimination (Section~\ref{sec:introfairness}), discuss notions of fairness proposed in the computer science literature (Section~\ref{sec:notionsfairness}), and explore some of the underlying causes of unfair predictions (Section~\ref{sec:biases}).

\section{Algorithmic Fairness}
\label{sec:introfairness}
Machine learning applications often make predictions about people. For example, algorithmic systems may be used to decide whether a resume makes it through the first selection round, judge the severity of a medical condition, or determine whether somebody will receive a loan. Since these systems are usually trained on massive amounts of data, they have the potential to be more consistent than human decision-makers with varying levels of experience. For example, consider a resume screening process. In a non-automated scenario, the likelihood to get through the resume selection round can depend on the personal beliefs of the recruiter who happens to judge your resume. On the other hand, the predictions of an algorithmic resume screening system can be learned from the collective judgement of many different recruiters.

However, the workings of a machine learning model heavily depend on how the machine learning task is formulated and which data is used to train the model. Consequently, prejudices against particular groups can seep into the model in each step of the development process. For example, if in the past a company has hired more men than women, this will be reflected in the training data. The machine learning model is likely to pick up this pattern. In fact, this is precisely what happened when Amazon tried to train a resume screening model\footnote{\url{https://www.reuters.com/article/us-amazon-com-jobs-automation-insight-idUSKCN1MK08G}}. The model did not explicitly take the applicant's gender into account. However, it turned out that the model penalized resumes that included terms that suggested that the applicant was female. For example, resumes that included the word ``women's" (e.g., in ``women’s chess club captain") were less likely to be selected. 

Notably, the characteristics that could potentially make algorithmic systems desirable over human-decision making, also amplify fairness related risks. One prejudiced recruiter can judge a few dozen resumes each day, but an algorithmic system can process thousands of resumes in the blink of an eye. If an algorithmic system is biased in any way, harmful consequences will be structural and can occur at an exceptionally large scale. 

Even in applications where predictions do not directly consider individuals, people can be unfairly impacted \citep{Barocas2019}. For example, a machine learning model that predicts the future value of houses can influence the actual sale prices. If some neighborhoods receive much lower house price predictions than others, this may disproportionately affect some groups over others.\\ 

\noindent Discrimination and bias of algorithmic systems is not a new problem. Well over two decades ago, \citet{Friedman1996} analyzed fairness of computer systems. However, with the increasing use of algorithmic systems, it has become clear that the issue is far from solved. Researchers from a range of disciplines have started working on unraveling the mechanisms in which algorithmic systems can undermine fairness and how these risks can be mitigated. This has given rise to the research field of \gls{algorithmicfairness}: the idea that algorithmic systems should behave or treat people fairly, i.e., without discrimination on the grounds of \glspl{sensitivecharacteristic} such as age, sex, disability, ethnic or racial origin, religion or belief, or sexual orientation.

In this definition, a \gls{sensitivecharacteristic} refers to a characteristic of an individual such that any decisions based on this characteristic are considered undesirable from an ethical or legal point of view. Note that our definition of \gls{algorithmicfairness} is very broad. This is intentional. The concept is applicable to all types of algorithmic systems, including different flavors of artificial intelligence (e.g., symbolic approaches, expert systems, and machine learning), but also simple rule-based systems. In this introduction to \gls{algorithmicfairness}, we will limit the discussion mostly to fairness of machine learning-based systems. 

\subsection{Types of Harm}
The exact meaning of ``behaving or treating people fairly" depends heavily on the sociotechnical context of the algorithmic system. There are several different ways in which algorithmic systems can disregard fairness.
\begin{itemize}
    \item \Gls{allocationharm} can be defined as an unfair allocation of opportunities, resources, or information \citep{Madaio2020}. In our resume selection example, \gls{allocationharm} occurs when some groups are selected less often than others, e.g., the algorithm selects men more often than women.
    \item {\Gls{qualityofserviceharm}} occurs when a system disproportionately fails for certain (groups of) people \citep{Madaio2020}. For example, a facial recognition system may misclassify black women at a higher rate than white men \citep{buolamwini2018} and a speech recognition system may not work well for users whose disability impacts their clarity of speech \citep{Guo2019}. In our resume selection example, \gls{qualityofserviceharm} occurs when some groups are more often wrongly rejected than others; e.g., qualified women are selected at lower rates than qualified men.
    \item \Gls{stereotypingharm} occurs when a system reinforces undesirable and unfair societal stereotypes \citep{Madaio2020}. \Glspl{stereotypingharm} are particularly prevalent in natural language processing and computer vision systems, as societal stereotypes are often deeply embedded in text corpora and image labels. For example, an image search for ``CEO" may primarily show photos of white men.
    \item \Gls{denigrationharm} refers to situations in which algorithmic systems are actively derogatory or offensive \citep{Madaio2020}. For example, an automated tagging system may misclassify people as gorillas\footnote{\url{https://www.theverge.com/2015/7/1/8880363/google-apologizes-photos-app-tags-two-black-people-gorillas}} and a chat bot might start using derogatory slurs\footnote{\url{https://fortune.com/2020/09/29/artificial-intelligence-openai-gpt3-toxic/}}.
    \item \Gls{representationharm} occurs when the development and usage of algorithmic systems over- or under-represents certain groups of people \citep{Madaio2020}. For example, some racial groups may be overly scrutinized during welfare fraud investigations or neighborhoods with a high elderly population may be ignored because data on disturbances in the public space (such as potholes\footnote{\url{https://hbr.org/2013/04/the-hidden-biases-in-big-data}}) is collected using a smartphone app. \Gls{representationharm} can be connected to \gls{allocationharm} and \gls{qualityofserviceharm}. However, a lack of diversity by itself can already be considered a violation of fairness. Moreover, \gls{representationharm} can already occur even before the algorithmic system makes a prediction, which makes it important to consider from the start.
    \item \Gls{proceduralharm} occurs when decisions are made in a way that violates social norms \citep[see e.g.,][]{Rudin2018a}. For example, penalizing a job applicant for having more experience can be considered a form of \gls{proceduralharm}. \Gls{proceduralharm} is not limited to the prediction-generating mechanisms of the model itself, but can also be extended to the development and usage of the system. For example, is it communicated clearly that an algorithmic decision is made? Do data subjects receive a meaningful justification? Is it possible to appeal a decision? This form of \gls{proceduralharm} is closely related to algorithmic accountablity.
\end{itemize}
Note that these types of harm are not mutually exclusive and that this list is not complete -- there may be other context and application specific harms.

\section{Notions of Algorithmic Fairness} 
\label{sec:notionsfairness}
With a rising interest in fairer machine learning systems, different notions of \gls{algorithmicfairness} have been put forward in the computer science literature. Most of these notions focus on \gls{allocationharm} and \gls{qualityofserviceharm}. Considerably fewer studies consider \gls{stereotypingharm}, \gls{denigrationharm}, and \gls{proceduralharm}. Generally speaking, we can distinguish two lines of work: \gls{groupfairness} and \gls{individualfairness}.

\subsection{Group Fairness}
A straightforward way to approach fairness is to consider whether some groups are treated worse, on average, than other groups. \Gls{groupfairness} is a notion of fairness that requires group statistics to be equal across (sub)groups defined by \glspl{sensitivecharacteristic}. It is sometimes referred to as \gls{statisticalfairness}. \Gls{groupfairness} is an intuitive notion of fairness that is relatively easy to put into operation. For example, \gls{allocationharm} can be quantified as a \gls{groupfairness} metric by comparing a classifier's selection rates across different groups. We can quantify \gls{qualityofserviceharm} and \gls{representationharm} in a similar way.

\subsubsection{No Fairness Through Unawareness}
At this point, you may wonder: if we do not want to discriminate against certain groups, why don't we just remove the \gls{sensitivefeature}, i.e., the feature that represents a \gls{sensitivecharacteristic}, from the data set? Unfortunately, it is not that simple. To see why, it can help to distinguish between \gls{directdiscrimination} and \gls{indirectdiscrimination}.

In European Union law, \gls{directdiscrimination} refers to cases where (groups of) individuals are treated less favorably based directly on their membership of a protected-by-law group. In the United States, this is also referred to as disparate treatment. An example of \gls{directdiscrimination} is when a person is denied service in a restaurant based on their race. In the context of an algorithmic system, \gls{directdiscrimination} could occur when a machine learning model explicitly uses a \gls{sensitivefeature} to make a prediction. Following this definition of fairness, removing the \gls{sensitivefeature} will prevent discrimination.

\Gls{indirectdiscrimination} (known as disparate impact in United States labor law), refers to cases where groups or individuals are treated less favorably based on rules that seem neutral, but, as a side effect, disadvantage a protected group. A classic example of \gls{indirectdiscrimination} is \gls{redlining}. This refers to a practice in the United States where people were systematically denied services based on their postal code. Neighborhoods that were deemed ``too risky" were outlined on the map in the color red, hence the name \gls{redlining}. Although postal code may appear to be a neutral feature, it is highly correlated with ethnicity. As services were mostly denied in older, predominantly black neighborhoods, black people were indirectly discriminated by this policy. Another example can be found in loan applications. Imagine we want to avoid \gls{allocationharm} across genders. We decide to exclude the feature that represents gender from our data set, to avoid any \gls{directdiscrimination}. However, if we do include occupation, an attribute which is highly gendered in many societies, the model can still identify historical patterns of gender bias. As such, occupation acts as a \gls{proxyvariable} for gender: it is a variable that is highly associated with another variable. \Glspl{proxyvariable} often act as a simple \gls{measurementmodel} for an abstract \gls{unobservableconstruct} or a variable that is otherwise hard to measure directly, e.g., due to practical constraints. For example, the ``quality" of a sales employee is an \gls{unobservableconstruct} that is impossible to measure directly. However, we can measure sales figures or customer satisfaction ratings, which can give an indication of the overall quality of an employee. In the case of \gls{indirectdiscrimination}, variables included in our model may unintentionally act as a \gls{proxyvariable} for a sensitive characteristic. 

This is not just a hypothetical problem. Machine learning algorithms are specifically designed to identify relationships between features in a data set. Hence, if discriminatory patterns exist within the data, it is very likely that a machine learning model will replicate it. Removing all possible \glspl{proxyvariable} is usually not a viable approach. First of all, it is not always possible to anticipate the patterns through which the \gls{sensitivefeature} can be approximated by the model. Several features that are slightly predictive of the \gls{sensitivefeature} might, taken together, be an accurate predictor of the \gls{sensitivefeature}. Second, apart from their relation with the \gls{sensitivefeature}, \glspl{proxyvariable} may provide information that is predictive of the target feature. Removing all features that are slightly related to the \gls{sensitivefeature} could therefore substantially reduce the predictive performance of the model.

Clearly, removing \glspl{sensitivefeature} is unlikely to prevent \gls{allocationharm}, which can still occur in the form of \gls{indirectdiscrimination}. Similarly, patterns of stereotyping and denigration can be deeply embedded in the data and removing the \glspl{sensitivefeature} will not remove these patterns. Additionally, \gls{qualityofserviceharm}, \gls{representationharm}, and \gls{denigrationharm} can be caused by a lack of information on minority groups, which is not solved by removing the \gls{sensitivefeature} either. To conclude, removing \glspl{sensitivefeature} is only helpful in achieving a very narrow definition of fairness. The practical consequence is that it is unlikely that this approach will prevent real-world harms.

\subsubsection{Conditional Group Fairness}
\label{sec:conditionalgroupfairness}
In EU and US law, \gls{indirectdiscrimination} in employment may not be unlawful if it is justified by a so-called ``legitimate aim". Examples of legal justifications for discrimination are genuine occupational requirement and business necessity. For example, a film producer is allowed to hire only male actors to play a male role, as this is considered a genuine occupational requirement. Apart from employment, there may be characteristics that, from an ethical perspective, legitimize differences between groups. Loosely inspired by these legal imperatives, \citet{Kamiran2013} put forward a notion of fairness that we will refer to as \gls{conditionalgroupfairness}. This is a variant of \gls{groupfairness} that allows for differences between groups, if these differences are explained by a legitimate feature that can be justified by ethics and/or law.

\Gls{conditionalgroupfairness} is best illustrated by an example. Imagine a scenario in which women have a lower income, on average, than men. This may imply that women are discriminated against. However, in our scenario many women work fewer hours than men. The observed \gls{indirectdiscrimination} can therefore be at least partly explained by the lower number of working hours. Consequently, equalizing income between men and women would mean that women are paid more per hour than men. If we believe unequal hourly wages to be unfair, we can instead equalize income only between women and men who work similar hours. In other words, we minimize the difference that is still present after conditioning on working hours. \Gls{conditionalgroupfairness} is particularly relevant considering \textit{Simpson's paradox}. This paradox states that if a correlation occurs in several different groups, it may disappear or even reverse when the groups are aggregated (see Figure~\ref{fig:simpsonsparadox}). In Example~\ref{ex:simpsonsparadox}, we see that an analysis that does not consider all relevant characteristics might suggest discrimination in situations that would be considered morally acceptable if all information was known.

\def\sc{0.6}
\tikzstyle{anode}=[circle, fill=white!80!black, draw=black, minimum size=\sc*8, thick, inner sep=1pt]
\tikzstyle{bnode}=[circle, fill=black, draw=black, minimum size=\sc*8, thick, inner sep=1pt]
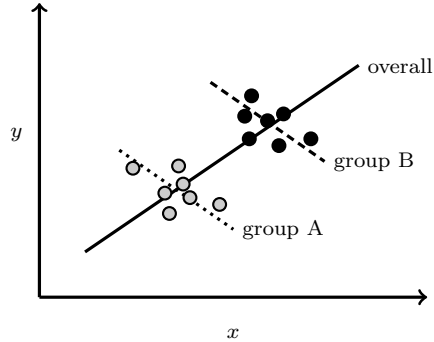
\begin{figure}[!ht]
    \footnotesize
    \centering
    \begin{tikzpicture}[very thick]
        % axes
        \draw[->] (-0.5*\sc,-0.5*\sc) -- (8*\sc,-0.5*\sc) node[below, midway, shift={(0*\sc,-0.5*\sc)}] {$x$};
        \draw[->] (-0.5*\sc,-0.5*\sc) -- (-0.5*\sc,6*\sc) node[above, midway, shift={(-0.5*\sc,0*\sc)}] {$y$};
        
        % plots
        \draw [solid] plot [] coordinates {(0.5*\sc, 0.5*\sc) (6.5*\sc,4.625*\sc)};
        \draw [dotted] plot [] coordinates {(1.25*\sc, 2.75*\sc) (3.75*\sc,1*\sc)};
        \draw [densely dashed] plot [] coordinates {(3.25*\sc, 4.25*\sc) (5.75*\sc,2.5*\sc)};
        
        % annotation
        \draw (6.5*\sc,4.625*\sc) node[anchor=west] {overall};
        \draw (3.75*\sc,1*\sc) node[anchor=west] {group A};
        \draw (5.75*\sc,2.5*\sc) node[anchor=west] {group B};
        
        % nodes group A
        \node[anode] at (1.55*\sc,2.35*\sc) {};
        \node[anode] at (2.55*\sc,2.40*\sc) {};
        \node[anode] at (2.35*\sc,1.35*\sc) {};
        \node[anode] at (3.45*\sc,1.55*\sc) {};
        \node[anode] at (2.65*\sc,2*\sc) {};
        \node[anode] at (2.25*\sc,1.80*\sc) {};
        \node[anode] at (2.80*\sc,1.70*\sc) {};

        % nodes group B
        \node[bnode] at (4.15*\sc,3.95*\sc) {};
        \node[bnode] at (4.85*\sc,3.55*\sc) {};
        \node[bnode] at (4.75*\sc,2.85*\sc) {};
        \node[bnode] at (5.45*\sc,3.00*\sc) {};
        \node[bnode] at (4.10*\sc,3.00*\sc) {};
        \node[bnode] at (4.00*\sc,3.50*\sc) {};
        \node[bnode] at (4.50*\sc,3.40*\sc) {};
    \end{tikzpicture}
    \caption{Visualization of Simpson's paradox. The correlation in both group $A$ and group $B$ is negative, whereas the overall regression line has a positive slope.}
    \label{fig:simpsonsparadox}
\end{figure}

\begin{example}\textit{Simpson's Paradox: Berkeley University Admissions}.
\label{ex:simpsonsparadox}
A classic example of Simpson's paradox is Berkeley's university admissions in 1973. 

When considering all programs together, women were accepted less often than men, implying a gender bias. However, it turned out that women at Berkeley often apply for competitive programs with a relatively low acceptance rate. As a result, the overall acceptance rate of women in the aggregated data was lower -- even though the acceptance rate of women \textit{within} each program was higher than the acceptance rate of men. Hence, if the admission's office would have tried to equalize the overall acceptance rate between men and women, men would have received an even lower acceptance rate.
\end{example}

\subsection{Individual Fairness}
So far, we have considered fairness from a group perspective. \citet{Dwork2012} put forward an alternative notion of fairness: \gls{individualfairness}. In this line of work, an algorithmic system is deemed fair if similar cases are treated similarly, irrespective of sensitive group membership. 

This notion of fairness resembles \gls{situationtesting}, a traditional discrimination auditing approach. In this approach, pairs of individuals that are equal, except for their membership of a protected group, are put in the same situation. If the member of a protected group is treated less favorably, this is regarded as discrimination. For example, an auditor may send out pairs of fictional resumes that are identical apart from the gender of the fictional applicant. Notions of \gls{individualfairness} can be used to quantify \gls{allocationharm} and \gls{qualityofserviceharm}, by comparing the prediction or performance of the model between pairs of similar instances.

Putting \gls{individualfairness} to practice can be challenging, as it requires a notion of similarity. A naive approach would be to simply change an instance's \gls{sensitivefeature} and see how this affects the machine learning model's prediction. However, such an approach has limitations similar to removing the \gls{sensitivefeature} from the data set. In particular, there may exist a relationship between the \gls{sensitivefeature} and other features, which mean that the perturbed instances would hardly ever (or never) occur in reality. When we limit ourselves to comparing cases that actually occur in reality, we require a more sophisticated notion of similarity. How can we incorporate the relative importance of features? How can we assess similarity across multiple features with different domains? How similar is similar enough? Defining a similarity metric is a highly non-trivial task. 

\subsubsection{Counterfactual Fairness}
The notions of \gls{groupfairness} and \gls{individualfairness} discussed so far approach fairness primarily from a statistical point of view. The notions acknowledge that there exist relationships between features, but the associated challenges are approached from a correlation-based perspective. In a different line of work, \citet{Kusner2017} propose to take an explicit causal approach that leverages causal models.

Causal models are mathematical models, typically graphs, that represent causal relationships between different features. Causal models cannot be learned from data alone: we cannot know from observations whether the crowing of the rooster caused the sun to rise, or the other way around. Instead, the structure of a causal model is based on assumptions about causal relationships, grounded in our understanding of how the world works. Observational data can be used to learn the strength of relationships. An interesting feature of causal models is that they allow for answering counterfactual questions: \textit{``what would have happened if..."?}

In the context of fairness, a causal model can help to determine what the model would have predicted if the individual had belonged to another group. \Gls{counterfactualfairness} is a notion of fairness that requires that the treatment of an individual in the actual world is the same as the treatment of the individual in a counterfactual world where the individual belonged to a different sensitive group. Similar to \gls{individualfairness}, \gls{counterfactualfairness} considers fairness from an individual perspective. However, rather than considering a ``close-enough" world by means of a similarity metric, \gls{counterfactualfairness} criteria are informed by counterfactual probabilities inferred from a causal model. By approaching \gls{individualfairness} through a causal lens, assumptions about relationships between features can be made explicit. This alleviates some of the issues involved with defining a similarity metric. The challenge, of course, moves from defining a similarity metric to defining a causal model. As mentioned before, the structure of a causal model cannot be identified from data alone. Instead, it requires a deep understanding of the data generating process. Different assumptions about this process can lead to vastly different models and, consequently, different conclusions regarding fairness. A more philosophical challenge of notions of \gls{individualfairness} is that sensitive group membership is often an important part of somebody's identity. Consequently, one may wonder what it means for an individual to ``belong to a different sensitive group".

\section{Biases as Sources of Unfairness}
\label{sec:biases}
% \todo{Add visualization that shows in which parts of the process biases can seep into the system}
In the previous section we have already touched upon some of the mechanisms through which social bias may be replicated from historical decision-making. However, this is not the full story. Algorithmic systems are an accumulation of design choices that embed the developers' explicit and implicit value judgements into the system \citep{Wieringa2020}. Consequently, biases can seep into the system in many different places of the development process. In this section, we will explore biases as sources of unfairness in different parts of the machine learning development process. 

\subsection{Types of Bias}
But first, let us define more precisely what we mean by bias. Generally speaking, \gls{bias} is a systematic and disproportionate tendency towards something. The word bias can be used to refer to many different things, ranging from social biases related to prejudice, to statistical biases that are of a technical nature.

\subsubsection{Social Bias}
In everyday language, bias often considers prejudice against a person or a group, which we will refer to as \gls{socialbias}. Social bias is a form of \gls{cognitivebias}: a systematic error in rational thinking that can affect judgment and decision-making. Most \glspl{cognitivebias}es are a result of the limitations of information-processing capabilities of the human brain. From an evolutionary perspective, these biases are useful because they allow people to make quick decisions in critical scenarios. As a hunter-gatherer, you would probably rather be safe than sorry when encountering an unknown group of other humans. However, these shortcuts often come at the cost of the quality of decisions. In particular, stereotypes formed by social bias can be overgeneralized and inaccurate, especially on an individual level. When people act on social biases, they can result in discriminatory practices. It is important to realize that everybody has some degree of conscious or subconscious social bias. Awareness of \glspl{cognitivebias}es, including social bias, can help to signal and mitigate their effects. In particular, a diverse team and critical self-reflection can help to signal social biases and avoid acting on them.

\subsubsection{Statistical Bias}
In statistics, bias refers to a systematic error in the estimation of parameters or variables. \Gls{statisticalbias} can be the result of data collection practices that compromise the accuracy of the estimate, such as a particular sampling procedure. This form of statistical bias can be rooted in \glspl{cognitivebias}es of the researcher or data subjects. Statistical bias can also refer to systematic errors caused by assumptions of the estimator. In the context of machine learning algorithms, this type of bias is often discussed in relation to the bias-variance trade-off. For example, some machine learning algorithms can only learn linear relationships between features, whereas the true underlying data distribution exhibits more complex relationships, resulting in an \gls{underfitting} model. \\ 

\noindent Although all these different types of bias can result in fairness related harms, most issues arise at the intersection of social bias and statistical bias. In the remainder of this section, we will dive more deeply into different types of biases at each stage of the development process. Note that this list is not exhaustive. Moreover, as we will see, biases are hard to precisely dissect and often overlap. In practice, it is usually difficult if not impossible to know exactly which biases are at play. Luckily, a thoughtful development process that leaves room for self-reflection goes a long way in mitigating harms.

\subsection{Problem Understanding: Abstraction Traps}
\label{sec:abstractiontraps}
Translating a real-world problem into a machine learning task can be difficult. By definition, a model is a simplification of reality. A data scientist's task is to decide which elements of the real world need to be included in the model and which elements will be left out of scope. To this end, some amount of abstraction is required. This involves removing details to focus attention on general patterns. The impact of your system, both positive and negative, highly depends on how you define the machine learning task. By abstracting away the context surrounding a model and its inputs and outputs, you may accidentally abstract away some of the consequences as well.

A mismatch between the machine learning task and the real-world context is referred to as an \gls{abstractiontrap} \citep{Selbst2019}. \Glspl{abstractiontrap} can amplify harmful consequences of your system, particularly those related to fairness. We will now discuss five \glspl{abstractiontrap}: the \gls{framingtrap}, \gls{portabilitytrap}, \gls{formalismtrap}, \gls{rippleffecttrap}, and \gls{solutionismtrap}.

\subsubsection{The Framing Trap} 
Machine learning models hardly ever operate in isolation. A decision-making process may incorporate (multiple) other machine learning models or human decision-makers. The \gls{framingtrap} considers the failure to model the relevant aspects of the larger system your machine learning model is a part of. 

For example, consider a scenario in which judges need to decide whether a defendant is detained. To assist them in their decision making process, they may be provided with a machine learning model that predicts the risk of recidivism; i.e., the risk that the defendant will re-offend. Notably, the final decision of the judge determines the real-world consequences, not the model's prediction. Hence, if fairness is a requirement, it is not sufficient to consider the output of the model; you also need to consider how the predictions are used by the judges. 

In many real-world systems, several machine learning models are deployed at the same time at different points in the decision-making process. Unfortunately, a system of components that seem fair in isolation do not automatically imply a fair system, i.e. \textit{Fair + Fair $\neq$ Fair} \citep{Dwork2020}.

To avoid the \gls{framingtrap}, we need to ensure that the way we frame our problem and evaluate our solution includes all relevant components and actors of the sociotechnical system.

\subsubsection{The Portability Trap} 
A system that is carefully designed for a particular context cannot always be directly applied in a different context. Taking an existing solution and applying it in a different situation without taking into account the differences between the two contexts is known as the \gls{portabilitytrap}. A shift in domain, geographical location, time, or even the nature of the decision-making process all impact the suitability of a system. For example, a voice recognition system optimized for speakers with an Australian accent may fail horribly when deployed in the United States. Similarly, the expectations of a good manager have changed considerably in the past few decades, including a stronger need for soft skills. A model trained on annual reviews in the 1960's will likely not be suitable to make predictions for current managers. The \gls{portabilitytrap} goes beyond performance issues due to differences in the data distribution. It also considers differences in social norms and actors. For example, a chat bot optimized to formulate snarky replies may be considered funny on a gaming platform, but inappropriate or even offensive in a more formal context, such as a website for loan applications. To avoid falling into the \gls{portabilitytrap}, we need to consider whether our problem understanding adequately models the social and technical requirements of the actual deployment context.

\subsubsection{The Formalism Trap}
In order to use machine learning, you need to formulate your problem in a way that a mathematical algorithm can understand. This is not a straightforward task: there are usually many different ways to measure something. Some may be more appropriate than others. You fall into the \gls{formalismtrap} when your formalization does not adequately take into account the context in which your model will be used. For example, machine learning problem formulations often simplify the decision space to a very limited set of actions \citep{Mitchell2018}. In lending, the decision space of the machine learning model may consist of two options: reject or accept. In reality, there may be many more actions available, such as recommending a different type of loan.

The \gls{formalismtrap} is closely related to the statistical concept of \gls{constructvalidity}: how well does the formalization measure the construct of interest? Business objectives often involve constructs such as ``employee quality" or ``creditworthiness" that cannot be measured directly \citep{Jacobs2019}. In such cases, data scientists may use a \gls{proxyvariable} instead. Every variable in a data set is the result of a decision on how a particular construct can be measured into a computer readable scale. For example, Netflix has chosen to measure a viewers' quality judgments with likes, rather than the more commonly used 1 - 5 star rating \citep{Dobbe2018}.  

Not all variables measure the intended construct equally well. \Gls{constructvaliditybias} is a statistical bias that occurs when a variable does not accurately measure the construct it is supposed to measure. For example, income measures the construct socioeconomic status to some degree, but does not capture other factors such as wealth and education \citep{Jacobs2019}.

A mismatch between the choice of target variable and the actual construct of interest can be detrimental towards fairness goals. In particular, fairness concerns can arise when the measurement error introduced by the choice of formalization differs across groups. For example, you may be interested in predicting crime, but only have access to a subset of all criminal activity: arrest records. In societies where arrest records are the result of racially biased policing practices, the measurement error will differ across racial groups. Similarly, \citet{obermeyer2019} found that due to unequal access to healthcare, historically less money has been spent on caring for African-American patients compared to Caucasian patients. Consequently, a system that used healthcare costs as a proxy for true healthcare needs systematically underestimated the needs of African-American patients.

Issues of \gls{constructvalidity} are especially complex for social constructs such as race and gender. Almost paradoxically, measuring \glspl{sensitivecharacteristic} can introduce bias. In industry and academia, it is common to ``infer" these characteristics from observed data, such as facial analysis \citep{Jacobs2019}. This can be problematic because such approaches often fail to acknowledge that social constructs are inherently contextual, may change over time, and are multidimensional. For example, when talking about race, one may be referring to somebody's racial identity (i.e., self-identified race), their observed race (i.e., the race others believe them to be), their phenotype (i.e., racial appearance), or even their reflected race (i.e., the race they believe others assume them to be). Which dimension you measure will influence the conclusions you can draw (see \citet{Hanna2020} for a more detailed account).

To avoid falling into the \gls{formalismtrap}, data scientists should take into account whether the problem formulation handles understandings of (social) constructs in a way that matches the intended deployment context. To mitigate \gls{constructvaliditybias}, ideally multiple measures are collected, especially for complex constructs. As we will see, many of the biases that can occur in later steps of the development process can be traced back to the problem of \gls{constructvalidity}.

\subsubsection{The Ripple Effect Trap}
Introducing a machine learning model in a social context may affect the behavior of other actors in the system and, as a result, the context itself. This is known as the \gls{rippleffecttrap}. There are several ways in which a social context may change due to the introduction of a new technology. First, the introduction of a new technology might be used to argue for or reinforce power, which can change an organization's dynamics. For example, management may purchase software for monitoring workers, reinforcing the power relationship between management and subordinates. Second, the introduction of a prediction systems may cause reactivity behavior. For example, people might attempt to game an automated loan approval system by dishonestly filling out their data in the hope of a more favorable outcome. Third, a system that was developed for a particular use case, may be used in unintended, perhaps even adversarial, ways. To avoid falling into the \gls{rippleffecttrap}, it is important to consider whether the envisioned system changes the context in a predictable way.

\subsubsection{The Solutionism Trap}
The possible benefits that machine learning solutions can bring can be very exciting. Unfortunately, machine learning is not the answer to everything (\textit{what?!}). The belief that every problem has a technological solution is referred to as \gls{solutionism}. We fall into the \gls{solutionismtrap} when we fail to recognize the machine learning is not the right tool for the problem at hand.

The \gls{solutionismtrap} is closely related to the \gls{optimismbias}. This is a \gls{cognitivebias} that causes people to overestimate the likelihood of positive events and underestimate the likelihood of negative events. In the context of algorithmic systems, \gls{optimismbias} occurs when policy makers or developers are overly optimistic about a system's benefits, while underestimating its limitations and weaknesses. In particular, people might overestimate the objectiveness of data and algorithmic systems. If this happens, the system's goals, development, and outcomes might not be sufficiently scrutinized, which can result in systematic harms. 

There are several reasons why machine learning may not be the right tool to solve a problem. In some scenarios, it may not be possible to adequately model the context using automated data collection. For example, consider eligibility for social welfare benefits in the Netherlands. Although the criteria for eligibility are set in the law, some variables, e.g. living situation, are difficult to measure quantitatively. Moreover, the Dutch legal system contains the possibility to deviate from the criteria due to compelling personal circumstances. It is impossible to anticipate all context-dependent situations in advance. As a result, machine learning may not be the best tool for this job. In other scenarios, machine learning may be inappropriate because it lacks human connection. For example, consider a person who is hospitalized. In theory, it may be possible to develop a robot nurse who is perfectly capable of performing tasks such as inserting an IV or washing the patient. However, the patient may also value the genuine interest and concern of a nurse  -- in other words, a human connection, something a machine learning model cannot (or even should not) provide. Furthermore, there may be cases where machine learning is simply overkill. For example, you may wonder whether spending several months on optimizing a deep learning computer vision system to predict the dimensions of items in your online shop is a better approach than simply asking the person who puts the item on the website to fill out the dimensions.  

To avoid falling into the \gls{solutionismtrap}, it is useful to consider machine learning as a means to an end. In other words, rather than asking ``can we use machine learning", ask, ``how can we solve this problem?" and then consider machine learning as one of the options.

\subsection{Data Collection and Processing}
It may not come as a surprise that many fairness issues arise through biases in data collection and analysis. We will discuss three forms of bias that may be present in the data: \gls{historicalbias}, \gls{representationbias}, \gls{measurementbias}.

\subsubsection{{Historical bias}}
Social biases can be encoded in data. If not accounted for, a machine learning model will reproduce these biases, resulting in unfair outcomes. Generally speaking, \gls{historicalbias} comes in two flavors. 

Firstly, \gls{historicalbias} can arise due to social biases in human decision-making. This type of bias is particularly prevalent when target labels are based on human judgement. For example, if in the past more men have been hired than women, a model trained on historical decisions will likely reproduce this association. As discussed in the previous section, simply removing the \gls{sensitivefeature} is not likely to remove this type of bias due to associations between features. Note that this type of \gls{historicalbias} is a form of \gls{constructvaliditybias}: the historical hiring decisions are a biased proxy for actual suitability of the applicant. Similarly, inaccurate stereotypes can be embedded in texts, images, and annotations produced by people, resulting in systems that reinforce these stereotypes.

A second type of \gls{historicalbias} occurs when the data is a good representation of reality, but reality is biased. In our hiring example, the observed bias could be caused by actual differences in suitability, but these differences are in turn caused by structural inequalities in society. For example, people from lower socioeconomic backgrounds may have had fewer opportunities to get good education, making them less suitable for jobs where such education is required for job performance. Similarly, some stereotypes are accurate at an aggregate level (even if they can be very inaccurate at an individual level!). For example, in many societies female nurses still greatly outnumber male nurses.

Depending on your worldview, you might have a different definition of what is fair in each of these scenarios. In practice, it is usually impossible to distinguish between these two types of \gls{historicalbias} from observed data alone. Moreover, they can also occur simultaneously. Consequently, understanding \gls{historicalbias} and identifying a mitigation approach that is in line with your own moral values requires a deep understanding of the social context.

\subsubsection{{Representation bias}}
\Gls{representationbias} occurs when some groups are underrepresented in the data \citep{suresh2020}. A machine learning model might not generalize well for underrepresented groups, causing \gls{qualityofserviceharm}. \Gls{representationbias} is especially risky when the data distribution of minority groups differs substantially from the majority group (see also \gls{aggregationbias}). A well-known example of \gls{representationbias} was uncovered by \citet{buolamwini2018}. They found that the data sets that were used to train commercial facial recognition systems contained predominantly images of white men. Consequently, the models did not generalize well to people with dark skin, especially women.

\Gls{representationbias} is closely related to \gls{selectionbias}, a statistical bias that occurs when the data collection or selection results in a non-random sample of the population. If not taken into account, conclusions regarding the studied effect may be wrong. For example, young healthy people may be more likely to volunteer for a vaccine trial than less healthy older people. As a result, conclusions about the side effects may not be representative for the whole population. Notably, \gls{representationbias} can occur even when a sample is truly random, as there may not be sufficient information available for minority groups. Moreover, \gls{representationbias} can be an issue in both training and testing data.

An underlying cause of \gls{representationbias} are blind spots of the collectors. For example, a data science team that consists solely of women is less likely to notice that men are not well represented in the data than a more diverse team. Additionally, some data is easier to get than others. For example, collecting data on the interests of young adults, who often spend hours each day scrolling through their social media feeds, is much easier compared to the interests elderly people who are generally not as active online. 

\Gls{representationbias} is relatively easy to solve by getting more data. Additionally, data scientists can leverage techniques that were designed to deal with sampling errors, such as weighting instances. However, these approaches first require you to identify cases in which \gls{representationbias} may occur. Therefore, a more durable solution is to invest in a diverse and inclusive development approach, which will help to avoid leaving groups out of the picture in the first place.

\subsubsection{{Measurement bias}}
\Gls{measurementbias} is a statistical bias that occurs when data contains systematic errors, due to data collection practices. \Gls{measurementbias} can be a cause of \gls{constructvaliditybias}. If systematic errors are correlated to \glspl{sensitivecharacteristic}, \gls{measurementbias} can become a source of unfairness.

\Gls{measurementbias} can occur when the method of observation results in systematic errors. First of all, the measurement process may be different across groups \citep{suresh2020}, due to a combination of social bias and \gls{confirmationbias}. \Gls{confirmationbias} is a \gls{cognitivebias} that refers to people's tendency to look for evidence of existing beliefs and disregard evidence that goes against it. In the context of data analysis, \gls{confirmationbias} can lead to cherry picking data that support a conclusion. As the famous quote by Ronald Coase says: ``\textit{if you torture the data long enough, it will confess to anything}." For example, a fraud analyst might overly scrutinize some groups over others. Higher rates of testing will result in more positives, confirming the analysts biased beliefs and skewing the observed base rates. If not accounted for, these skewed numbers will be reproduced by the machine learning model.

Another example of \gls{measurementbias} occurs when different observers interpret reality differently. In medical studies, for example, clinicians might arbitrarily round blood pressure readings up or down to the nearest whole number, depending on what they expect to see. In a machine learning context, this type of bias can occur when annotations reflect social biases, such as stereotypes, of the annotators or decision-makers.

\Gls{measurementbias} can also occur at the side of the data subject. Data subjects might behave differently because they are being observed, especially when data is collected from memory or through self-reporting. Survey responses may be incomplete or inconsistent because participants try to present themselves in a way that is socially desirable. For example, consider self-reported height, scraped from a dating website. In many cultures, tallness is seen as an attractive trait in men. Consequently, men may have exaggerated their height to appear more attractive, resulting in \gls{measurementbias}. Note that this is another example of how \gls{measurementbias} can be a threat to \gls{constructvalidity}.

\Gls{measurementbias} can be mitigated by high-quality data collection procedures. Note that it is not possible to identify cases of \gls{measurementbias} from observational data alone. For example, we cannot know the true underlying fraud rate if we only take into account data produced by a biased fraud detection approach. This highlights the importance of documenting the data collection procedure. % This does not only hold for data that was collected within your organization. Take into account possible \gls{measurementbias} when considering external data sources as well.

\subsection{Modeling}
Building a machine learning model includes many different choices, ranging from the class of machine learning algorithms that is considered to their hyperparameter settings. Different models may have different consequences related to fairness, depending on the task at hand.

\subsubsection{{Aggregation bias}}
\Gls{aggregationbias} occurs when a single model is used for groups that have distinct data distributions \citep{suresh2020}. If not accounted for, it may lead to a model that does not work well for any of the subgroups. For example, it is known that the relationship between hemoglobin levels (HbA1c) and blood glucose levels differs greatly across genders and ethnicities \citep{suresh2020}. If these differences are not taken into account, a model that only uses a single interpretation of HbA1c will likely not work well for any of these groups. In combination with \gls{representationbias}, it can lead to a model that only works well for the majority population. 

\Gls{aggregationbias} is related to the problem of \gls{underfitting}. Machine learning can be seen as a compression problem that produces a mapping between input features and output variable. Some information is inherently lost because of the chosen mapping \citep{Dobbe2018}. In particular, some model classes may not be able to adequately capture the different data distributions. Such an oversimplified model may come at the cost of predictive performance for minority groups, resulting in \gls{qualityofserviceharm}.

\subsubsection{{Omitted variable bias}}
\Gls{omittedvariablebias} is a statistical bias which occurs when one or more relevant features are left out of a linear regression model. Consequently, the model attributes the effects of the missing feature(s) to the included features, obscuring their true effects. \Gls{omittedvariablebias} was originally introduced in the context of statistical models that are used for causal inference. For example, consider a regression-based test for discrimination in loan applications. Imagine that loan officers consider payment history in their decision and that payment history correlates with race. If payment history is not recorded in the data, the results of the regression will attribute the effect of payment history to race, suggesting \gls{directdiscrimination} and potentially \gls{proceduralharm} when there is not \citep{jung2019omitted}. This bias can also be relevant for prediction tasks. In particular, excluding \glspl{sensitivefeature} from the data may obscure a model's indirect dependence on this feature, attributing their effects to other related features. This makes it more difficult to detect and account for existing \gls{historicalbias}.

\subsection{Evaluation}
During the evaluation stage, the final model is scrutinized in more detail. \Gls{evaluationbias} refers to the use of performance metrics and procedures that are not appropriate for the way in which the model will be used \citep{suresh2020}. \citet{Mitchell2018} identify several underlying assumptions of performance metrics. First, these metrics assume that individual decisions are independent of each other. Note how this assumption is grounded in {utilitarianism}, in which overall utility is expressed as the sum of individual utilities. In practice, however, the impact of a decision may not be independent across instances. For example, denying one family member a loan may impact another family member's ability to repay their own loan. Additionally, it is typically assumed that decisions are symmetrical, i.e. the impact of the outcome is equal across instances. Again, this often does not hold in practice. For example, a rejection of a job application can have a very different impact depending on whether that person is currently employed or unemployed.

\subsection{Deployment}
Once the system is deployed, it may be used, interpreted, or interacted with inappropriately, resulting in unfair outcomes \citep{Friedman1996}. The underlying cause of these outcomes is a mismatch between the system's design and the context in which it will be applied. Indeed, biases in deployment can often be attributed to \glspl{abstractiontrap} introduced in Section~\ref{sec:abstractiontraps}.

\paragraph{Usage}
The system may be used in a context for which it was not (properly) designed, in which case we fall into the \gls{portabilitytrap}. For example, a toxic language detection model trained on tweets may not be suitable for a platform such as TikTok, where the average user is much younger and may use different language (tone, words, etc.) than an average Twitter user. Note that this type of bias can also accrue over time due to changing populations and behaviors \citep{Mehrabi2019}, in which case it can be seen as a form of \gls{conceptdrift}.

\paragraph{Interpretation}
Interaction of stakeholders with the system can be a source of unfairness. A decision-maker may interpret the model's output differently for different groups, due to social bias and \gls{confirmationbias}. For example, a judge may weigh a high risk score more heavily for a black defendant compared to a white defendant, due to (unconscious) social bias. This bias, which can be attributed to falling into the \gls{framingtrap}, can be mitigated by taking into account stakeholder interactions during the system's design and evaluation.

\paragraph{Interaction}
In systems that learn from user interactions, users can introduce social bias. For example, consider a chat bot that learns dynamically. Without safeguards against toxicity, users might teach it to use obscene or otherwise offensive language, resulting in \gls{denigrationharm}. This type of bias can be avoided by putting checks in place to identify malicious intent towards the system.

\paragraph{\Gls{reinforcingfeedbackloop}}
Feedback mechanisms that amplify an effect are called \glspl{reinforcingfeedbackloop}. In the context of fairness, it refers to the amplification of existing (historical) biases when new data is collected based on the output of a biased model. 
\begin{example}
\label{ex:feedbackloop}
\textit{A Reinforcing Feedback Loop in Predictive Policing.}
Lets imagine there is a police station that is responsible for two neighborhoods, $A$ and $B$. Now lets imagine a predictive policing system that allocates police officers to the neighborhoods based on the predicted crime rate in each neighborhood. In this example, the true crime rates of the neighborhoods are equal. However, due to the randomness, we have collected slightly more crime data in neighborhood $A$ than than in neighborhood $B$ at the time the prediction model is trained. Consequently, the model predicts more crime in neighborhood $A$ than in neighborhood $B$. Based on this prediction, we send more police officers to neighborhood $A$. Consequently, more crime will be detected in neighborhood $A$ -- even though the true crime rates are the same. If we retrain our model on the newly collected crime data, even more police officers will be allocated to neighborhood $A$ and even more crime is detected. And so the feedback loop continues... 
\end{example}
A consequence of these feedback loops is that people can form erroneous beliefs based on the data. For example, after the introduction of the predictive policing system in Example~\ref{ex:feedbackloop}, police officers may believe that neighborhood $A$ truly has a bigger crime problem than neighborhood $B$. A failure to anticipate on feedback loops can be particularly risky for automated decision-making systems, in which bias can propagate quickly over time.

A specific instance of feedback loops that recommender systems may suffer from is \gls{popularitybias}. If people tend to click on highly ranked items more often, this can lead the algorithm to rank popular items even higher and disregard less popular items that may be just as valuable to the user. 

One way to investigate feedback loops is through simulation. Developing an accurate simulation of a sociotechnical system is difficult and requires a lot of domain expertise. Alternatively, we may borrow approaches from the field of system dynamics \citep{martin2020extending} and causal modeling.

\section{Discussion}
As we have seen in these notes, there is no standard notion of fairness in the literature on \gls{algorithmicfairness}. It should be stressed that this is a feature, not a bug. Different fairness criteria encode different value systems \citep{Hutchinson2019}, which can all be reasonable ethical stances. Rather than falling into the \glspl{abstractiontrap}, we need to acknowledge that fairness is an inherently contextual concept and no single definition of fairness will apply to all scenarios. Moreover, norms and laws shift over time and what is considered fair today may not be considered fair tomorrow. In practice, however, this means that it may be difficult to choose the ``right" notion of fairness. Identifying the ways in which the development or usage of your system might be harmful to vulnerable populations is often helpful in choosing an appropriate notion of fairness.

We have also seen that there are many different sources of unfairness, rooted in social biases we are not always aware of. Even a well-intended machine learning practitioner may inadvertently train a discriminatory model. Unraveling these biases is a complex task. In real-world settings, it is not possible to fully ``de-bias" a model. Instead, the goal of efforts in \gls{algorithmicfairness} is to mitigate potential harms as much as possible. As biases can seep in at every step of the way, mitigation requires careful attention throughout the entire development process.

% Finally, it is very difficult, if not impossible, to debate the fairness of a system without being transparent on how the system was developed and which trade-offs were made along the way. As such, we cannot emphasize enough how important it is to document the process carefully, especially for high-impact systems.

% ------------ Bibliography ------------ %
%
\bibliographystyle{abbrvnat}
\bibliography{references}

\setglossarystyle{index}
\printglossaries

\end{document}